\documentclass[fleqn,10pt]{wlscirep}

\title{Depth Compensated Spectral Domain Optical Coherence Tomography via Digital Compensation}

\author[1,*]{Ameneh Boroomand}
\author[2]{Bingyao Tan}
\author[1]{Mohammad Javad Shafiee}
\author[2,1]{Kostadinka Bizheva}
\author[1]{Alexander Wong}
\affil[1]{Dept. of Systems Design Engineering, University of Waterloo, Waterloo, Canada}
\affil[2]{Dept. of Physics and Astronomy, University of Waterloo, Waterloo, Ontario, Canada}

\affil[*]{aborooma@uwaterloo.ca}

\usepackage{float}%exact location
\usepackage{graphicx}
\usepackage{caption}
\usepackage{subcaption} % for subfigures

\begin{abstract}
Spectral Domain Optical Coherence Tomography (SD-OCT) is a well-known imaging modality which allows for \textit{in-vivo} visualization of the morphology of different biological tissues at cellular level resolutions. The overall SD-OCT imaging quality in terms of axial resolution and Signal-to-Noise Ratio (SNR) degrades with imaging depth, while the lateral resolution degrades with distance from the focal plane. This image quality degradation is due both to the design of the SD-OCT imaging system and the optical properties of the imaged object. Here, we present a novel Depth Compensated SD-OCT (DC-OCT) system that integrates a Depth Compensating Digital Signal Processing (DC-DSP) module to improve the overall imaging quality via digital compensation. The designed DC-DSP module can be integrated to any SD-OCT system and is able to simultaneously compensate for the depth-dependent loss of axial and lateral resolutions, depth-varying SNR, as well as sidelobe artifact for improved imaging quality. The integrated DC-DSP module is based on a unified Maximum a Posteriori (MAP) framework which incorporates a Stochastically Fully-connected Conditional Random Field (SF-CRF) model to produce tomograms with reduced speckle noise and artifact, as well as higher spatial resolution. The performance of our proposed DC-OCT system was tested on a USAF resolution target as well as different biological tissue samples, and the results demonstrate the potential of the proposed DC-OCT system for improving spatial resolution, contrast, and SNR.

\end{abstract}
\begin{document}

\flushbottom
\maketitle

%  Click the title above to edit the author information and abstract
\thispagestyle{empty}

\section*{Introduction}
Spectral Domain Optical Coherence Tomography (SD-OCT) is a well-established non-invasive interferometric optical imaging technique that provides micrometer scale spatial resolution in biological tissue over imaging depths of a few millimeters~\cite{fercher2003optical}. SD-OCT has found many different medical and industrial applications~\cite{tomlins2005theory}, where there is an on-going demand for observing and studying morphological details in the size range of a few micrometers.

There are limitations to the SD-OCT imaging quality in terms of spatial resolution~\cite{wojtkowski2004ultrahigh,fercher2003optical} and contrast, associated both with the optical design of the SD-OCT system and the structure of the imaged object.  The overall sensitivity (Signal-to-Noise Ratio (SNR)) of a SD-OCT system is basically dependent on several different factors in the SD-OCT system design including the camera spectral response, data acquisition rate, as well as the amount of power in the reference and sample arm~\cite{drexler2008optical}. The SNR roll-off within higher imaging depths happens due to the limited size and number of camera pixels combined with any chromatic and spherical aberrations in the spectrometer. The SNR roll-off typically is modeled using an exponential function within different imaging depths~\cite{drexler2008optical}.

In theory, the axial resolution of a SD-OCT system should be the same within all different imaging depths as it is only dependent on the central wavelength and spectral bandwidth of the light source~\cite{fercher2003optical}. However, in practice, the optical aberrations that happens in the spectrometer as well as the limited size and number of camera pixels allow for the cross-talk between neighboring pixels and for signals measured far away from the zero delay point of the scanning range. The occurrence of these cross-talks can change the spectrum and therefore degrade the effective axial resolution~\cite{fercher2003optical}. The other factors that affect the axial resolution include the light absorption and scattering that happens inside the imaged sample as they can alter the spectrum in the OCT sample arm and therefore degrade the effective axial resolution of acquired tomograms. These spectral changes are generally depth-dependent and impossible to quantify, as they are related to the morphology and reflectivity properties of the imaged object~\cite{wojtkowski2004ultrahigh}.

The theoretical lateral resolution of a SD-OCT system is primarily determined by the last imaging lens or microscope objective in the OCT imaging probe, as well as the central wavelength of the light source and therefore should be the same for all depth of imaging. However, in practice, the optical components of the OCT imaging probe introduce chromatic aberrations and have anti-reflectivity coatings that are spectrally dependent and therefore the lateral resolution degrades away from the focal plane due to defocus and aberrations. In addition, any depth-dependent absorption and scattering that happen inside the imaged object can change the spectrum and therefore the theoretical lateral resolution~\cite{fercher2003optical}.  Furthermore, sidelobe artifact is another factor which can degrade overall SD-OCT imaging quality. This artifact mainly happens due to the use of non-Gaussian optical sources in the design of SD-OCT system~\cite{piao2001cancellation,tripathi2002spectral}.  For better visualizing cellular level structures, fine features and micro-details of an imaged sample within higher imaging depths, it is essential to simultaneously compensate for all the degrading effects of the depth-dependent SNR roll-off, depth-varying axial and lateral resolutions as well as sidelobe artifact which all limit overall SD-OCT imaging quality.

Several image reconstruction methods have been previously proposed to compensate for the depth dependent SNR roll-off in an acquired tomogram to improve the overall SNR and consequently provide better visualization of the different micro-structures of the imaged sample ~\cite{liu2014noise,du2014speckle,jorgensen2007enhancing,wong2008perceptually}. Light sources with broader spectral bandwidth have also been utilized to overcome the physical limitation of the SD-OCT system in producing tomograms with the expected theoretical axial resolution~\cite{cheung2015high,fuchs2014towards}. Inverse scattering models~\cite{marks2006inverse} and different deconvolution algorithms~\cite{bousi2012axial,kulkarni1999image,liu2009deconvolution,schmitt1998restoration,hojjatoleslami2013image} have also been proposed to improve SD-OCT imaging quality by attenuating axial blurring that happens due to the OCT axial resolution degradation at higher imaging depths. In terms of lateral resolution, adaptive optics~\cite{hermann2004adaptive,wong2015vivo,miller2011adaptive}, axicon lenses~\cite{ding2002high,lee2008bessel}, dynamic focusing~\cite{murali2007dynamic} and focus tracking~\cite{cobb2005continuous} are the most well-known hardware approaches which are used to improve the lateral resolution beyond the Depth of Focus (DOF) range. Software replacement approaches are mainly limited to the lateral deconvolution methods~\cite{yasuno2006non,liu2009deconvolution,liu2011automatic}, inverse scattering models~\cite{ralston2006inverse} and also taking advantage of overlapping OCT measurements~\cite{bousi2012lateral} in reconstructing laterally resolved tomograms through an inverse model. In terms of sidelobe artifact removal, spectral shaping methods~\cite{tripathi2002spectral}, nonlinear deconvolution algorithms~\cite{piao2001cancellation} and linear filters~\cite{smith2001spectral} have been previously applied.  It can been seen that while different solutions have been proposed to tackle the degrading effects of the depth-dependent SNR roll-off, depth-varying axial and lateral resolutions as well as sidelobe artifact independently, strategies for addressing all of these issues in a unified framework has been largely unexplored.

Motivated by this, this study introduces a novel Depth Compensated SD-OCT (DC-OCT) system that integrates a Depth Compensating Digital Signal Processing (DC-DSP) module designed to concurrently compensate for depth-dependent SNR roll-off, the blurring effect due to the depth-varying axial and lateral resolutions, as well as sidelobe artifact to improve SD-OCT imaging quality. To the best of the authors' knowledge, such an approach to improving SD-OCT imaging quality has not been reported in previous studies.  The proposed DC-DSP module that can be integrated into any SD-OCT system is shown in Figure 1. The proposed DC-DSP module is based on a unified Maximum a Posteriori (MAP) framework that accounts for the depth-varying characteristics of the SD-OCT system to produce tomograms with higher spatial resolution, higher SNR, and reduced sidelobe artifact.  A novel aspect of the proposed work is the incorporation of the measured SNR curve as well as the calculated depth-varying axial and lateral Point Spread Functions (PSFs) of the utilized SD-OCT system into the modeling framework of the proposed DC-DSP module. This is helpful in compensating for the systematic limitation of SD-OCT system in providing high quality tomograms of different sample types. The other novel aspect of this work is in taking advantage of a recently-introduced random field modeling concept called the Stochastically Fully Connected Conditional Random Field (SF-CRF)~\cite{shafiee2014efficient,GSFCRF_2015_shafiee,DWI_2014_Shafiee} in the proposed DC-DSP module. The ability of the utilized SF-CRF model in modeling of the different interactions among the image pixels within the tomogram makes it suitable for mitigating the effect of speckle noise and artifact while improving the overall spatial resolution.

%----Figure 1---------------------
\begin{figure}[ht!]
	\centering
    \includegraphics[width=0.7\linewidth]{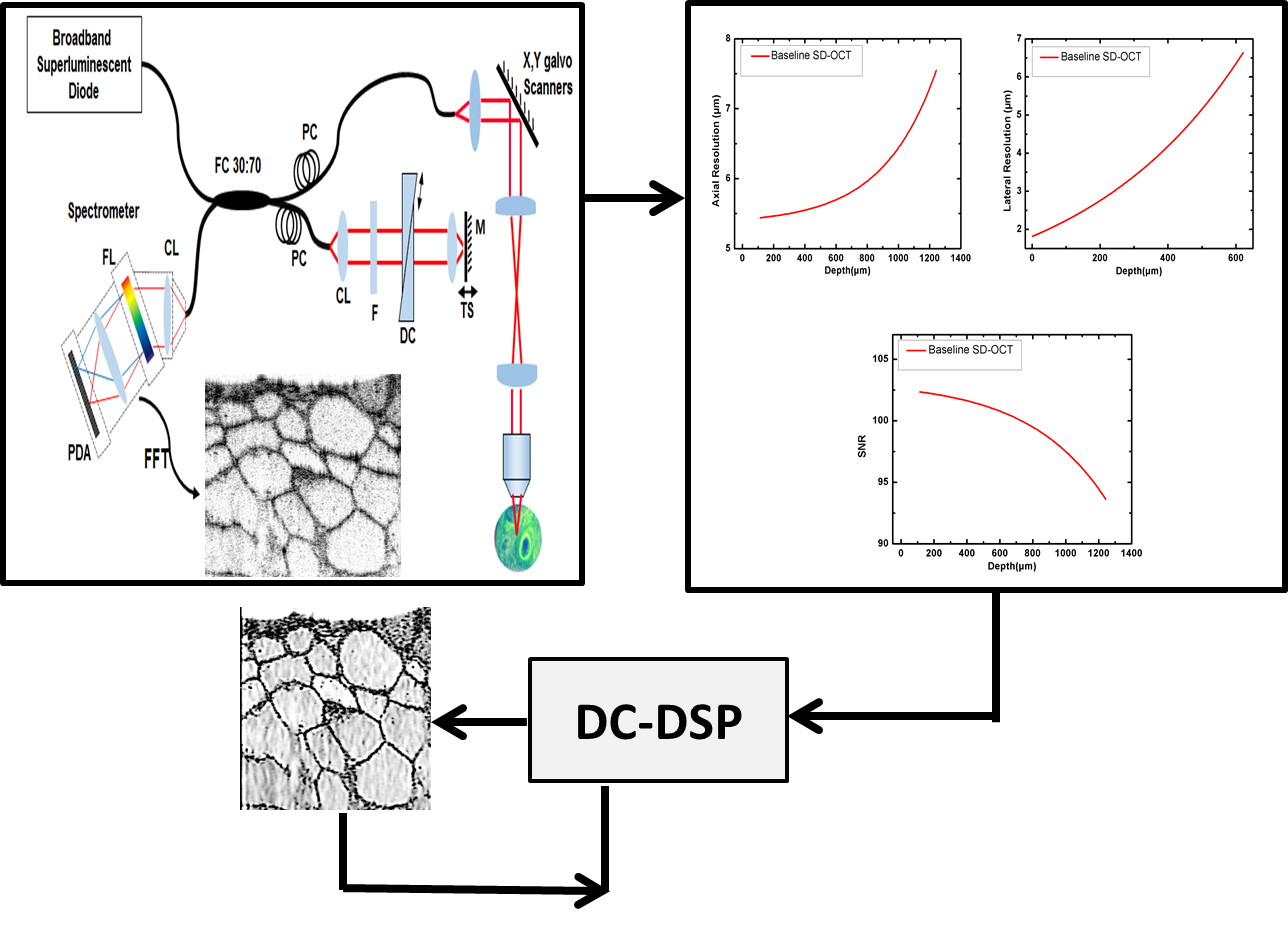}
	\caption{Schematic of proposed DC-OCT system. System components are; X-Y scanning mirrors, M-mirror, TS-translation stage, DC-dispersion compensation unit, F-neutral-density filter, CL-collimating lens, PC-polarization controllers, FC-fiber coupler, CL- collimating lens, FL-focusing lens, PDA-photodetector array}
	\label{fig:}
\end{figure}

\section*{Results}
In this section, we present the qualitative and quantitative results achieved using the proposed DC-OCT system for obtaining depth-compensated tomograms of several biological tissue samples. For comparison purposes, the performance of our proposed DC-OCT system is evaluated against the baseline SD-OCT system (which we will refer to as baseline-OCT) as well as the Maximum Likelihood SD-OCT system (ML-OCT)~\cite{hojjatoleslami2013image} which is the state of the art in obtaining depth-compensated tomograms. The tested ML-OCT system takes advantage of an extended Lucy-Richardson deconvolution strategy~\cite{singh2008adaptively} in order to digitally compensate for depth-varying lateral and axial resolution degradations.

To evaluate the ability of the proposed DC-OCT system to achieve high quality images of biological tissues, a thin slice of a grape and a thin slice of a lime were imaged using the SD-OCT setup, as shown in Figure 3(a,b). The theoretical axial and lateral resolutions are $2.49\mu m$ and $1.97 \mu m$ respectively for both grape and lime acquisitions using the baseline SD-OCT setup. Since both of the grape and lime samples consist of cells of different sizes with nearly spherical nuclei of several microns which can serve as point scatterers, these are proper samples for evaluating the efficacy of the tested systems in terms of improving the visualization of the different fine micro-structures, features and details of an imaged sample.

%-----Figure2----------------------
\begin{figure}[ht!]
	\centering
    \includegraphics[width=0.8\linewidth]{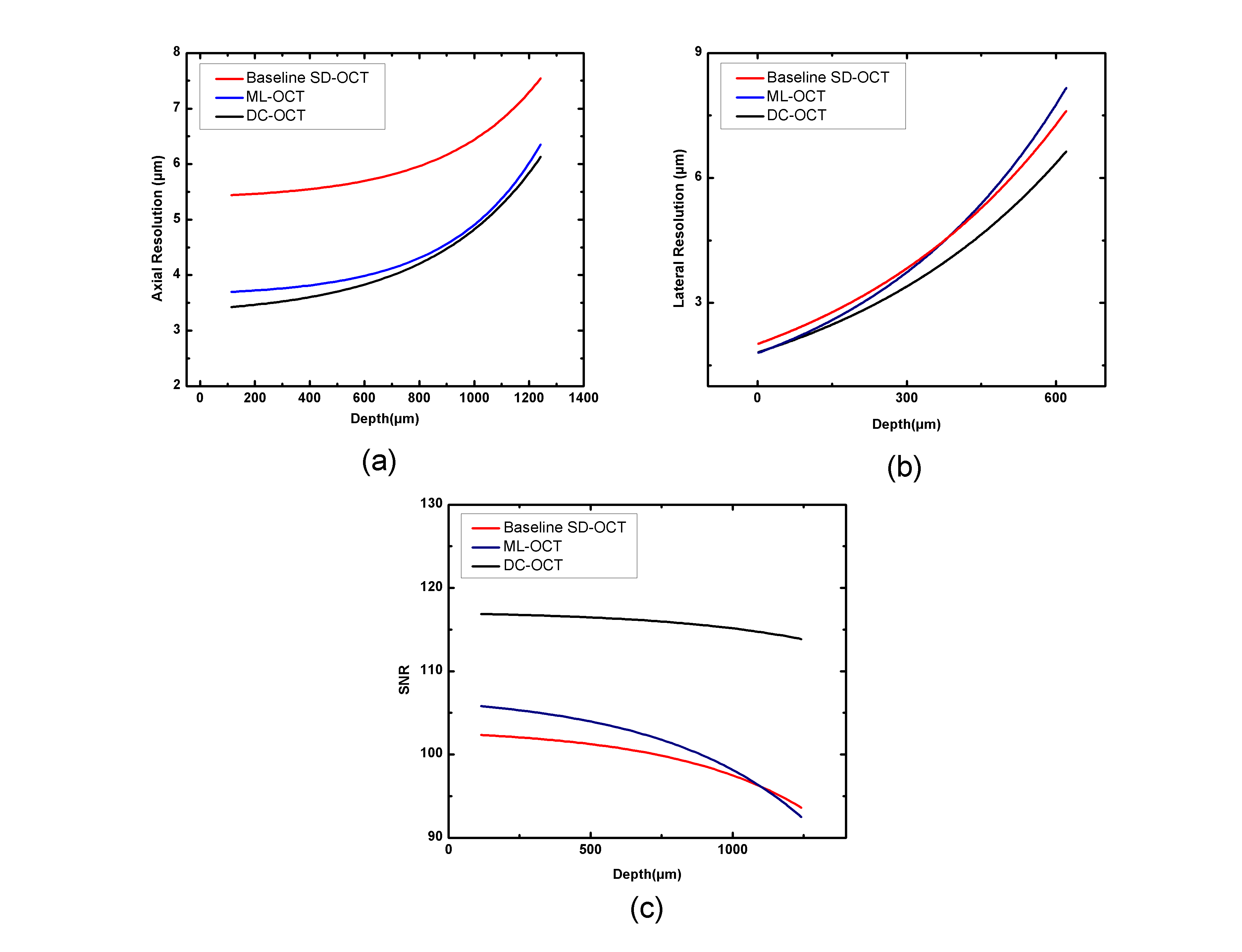}
	\caption{(a) Calculated axial resolution curves for acquired images of a standard U.S. Air Force resolution target using the baseline SD-OCT system (red curve), ML-OCT system (blue curve) and the proposed DC-OCT system (black curve). (b) Calculated lateral resolution curves for acquired images of a standard U.S. Air Force resolution target using the baseline SD-OCT system (red curve), ML-OCT system (blue curve) and proposed DC-OCT system (black curve).(c) Calculated SNR curves for acquired images of a standard U.S. Air Force resolution target using the baseline SD-OCT system (red curve), ML-OCT system (blue curve) and proposed DC-OCT system (black curve).}	
	\
\end{figure}

%----Figure 3----------------------
\begin{figure}[ht!]
\centering
    \includegraphics[width=0.7\linewidth]{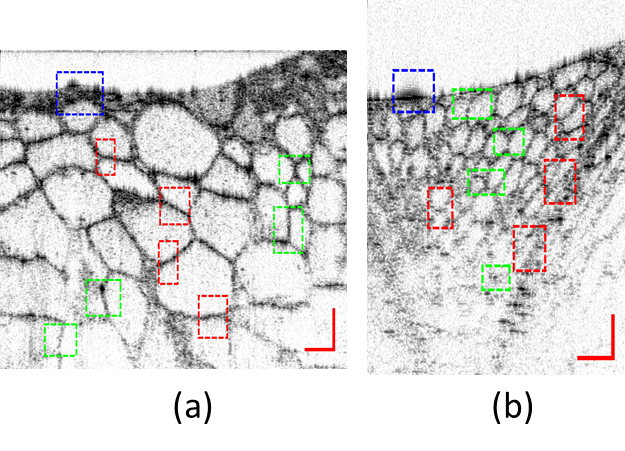}
\caption{(a) A slice of grape imaged using baseline SD-OCT system. (b) A slice of lime imaged using baseline SD-OCT system. The scale bars represent 75 $\mu m$ in both axial and lateral directions for the grape sample and 80 $\mu m$ in both axial and lateral directions for the lime sample. The blue color boxes in (a,b) represent a high reflective surface of the grape and lime samples which results in sidelobe artifact in acquired tomograms using the baseline SD-OCT system. The red color boxes marked in (a,b) denote to the areas of the grape and lime samples that are used for the axial PSF calculation in Figures 6 and 11. The green boxes in (a,b) represent the areas of the grape and lime samples that are used for the lateral PSF calculation in Figures 8 and 13.}
	\
\end{figure}

\subsection*{Qualitative Results}

A visual comparison of the results produced by the baseline-OCT system (Figure 4(a) and Figure 9(a)) with that produced by the proposed DC-OCT system (Figure 4(c) and Figure 9(c))  demonstrates the power of our proposed DC-OCT system in terms of spatial resolution enhancement at different imaging depths, as it can be clearly observed that small neighboring structures within the imaged samples are more distinguishable from each other when using the proposed DC-OCT system. As these results show, the boundaries of the sample cells are more clear and less blurry all along the different imaging depths while the nuclei of the small cells exhibit a more rounded shape, which is consistent with the actual shape of the nuclei for such cells, in the depth-compensated results produced using the DC-OCT system. Furthermore, the overall contrast is also enhanced by using the proposed DC-OCT system such that the small cell nuclei are more distinguished from the surrounding tissue in the results produced using the DC-OCT system (Figure 4(c) and Figure 9(c)) when compared to that produced using the baseline-OCT system (Figure 4(a) and Figure 9(a)). These results indicate the effectiveness of the produced DC-OCT system for compensating the different degrading effects of the SD-OCT system setup.

%-------------Figure 4-----------------------
\begin{figure}[ht!]
	\centering
    \includegraphics[width=0.9\linewidth]{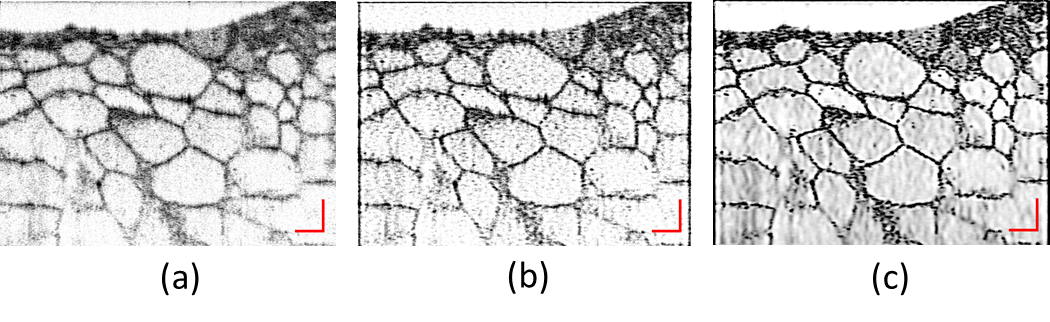}
	\caption{A slice of grape imaged using (a) baseline SD-OCT system, (b) ML-OCT system and (c) proposed DC-OCT system. The scale bars represent 75 $\mu m$ in both axial and lateral directions.}
	\
\end{figure}

In comparison with the results produced using the tested ML-OCT system, it can be observed that while the tested ML-OCT system (Figure 4(b) and Figure 9(b)) is able to produce results with sharper detail than the baseline-OCT system by compensating for the depth-varying axial and lateral resolution degradation, significant speckle noise artifact can be observed, thus leading to a reduction in SNR. In comparison, the results produced using the DC-OCT system is able to achieve noticeably improved detail visibility by mitigating speckle noise while improving axial and lateral resolutions. To have a better observation of the spatial resolution improvement, contrast enhancement, as well as SNR improvement, zoomed-in regions of interest around different cells boundaries and cell nuclei at various depths are chosen from the grape and lime samples (as highlighted by green and red color boxes in Figures 3(a,b)) and are shown in Figures 5 and 7 for the grape sample and in Figures 10 and 12 for the lime sample. As we can see in these zoomed-in regions, the cell boundaries appear sharper in both axial and lateral directions as well as more distinguishable from the surrounding regions in the results produced using the proposed DC-OCT system when compared to the baseline SD-OCT system and tested ML-OCT system. Furthermore, the small cell nuclei and other fine structures of the samples are more distinguishable from each other and there is a significant reduction in speckle noise in the results produced using the proposed DC-OCT system (for example, Figures 10 (d and f), Figures 10 (j and l) and also Figures 12(j and l)) when compared to the results produced using the ML-OCT system.

The blue color boxes marked on Figure 3(a,b) highlight regions of interest on the highly reflective surface of the samples that are significantly degraded by sidelobe artifact. It can be observed that the results produced by the baseline-OCT system (Figure 4(a) and Figure 9(a)) and the ML-OCT system (Figure 4(b) and Figure 9(b)) exhibit significant sidelobe artifact, which degrade the visibility of reflective structures and details in the imaged object. It can be further observed that in the results produced using the proposed DC-OCT system (Figure 4(c) and Figure 9(c)), the noticeable effect of sidelobe artifact is significantly reduced, which illustrates the effectiveness and importance of the proposed DC-OCT system in compensating for sidelobe effects.

\subsection*{Quantitative Results}

%------Figure 5---------------------------
\begin{figure}[ht!]
	\centering
    \includegraphics[width=0.5\linewidth]{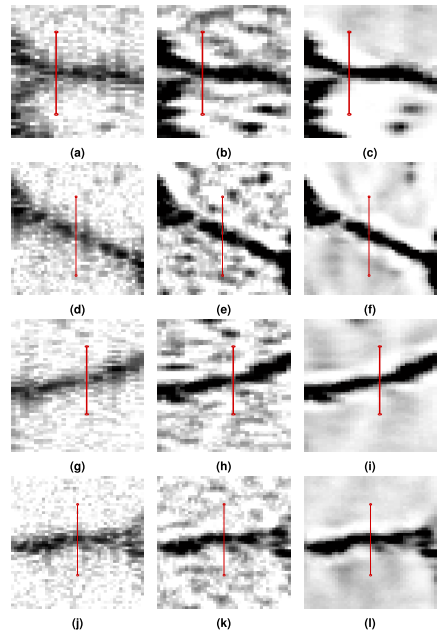}
	\caption{Enlarged regions representing different boundary cells of grape at various imaging depths as marked with red boxes in Figure 3(a) and for the baseline SD-OCT (a,d,g,j), ML-OCT (b,e,h,k) and proposed DC-OCT (c,f,i,l).}
     	
\end{figure}

%----Figure 6-------------------
\begin{figure}[ht]
	\centering	
    \includegraphics[width=0.8\linewidth]{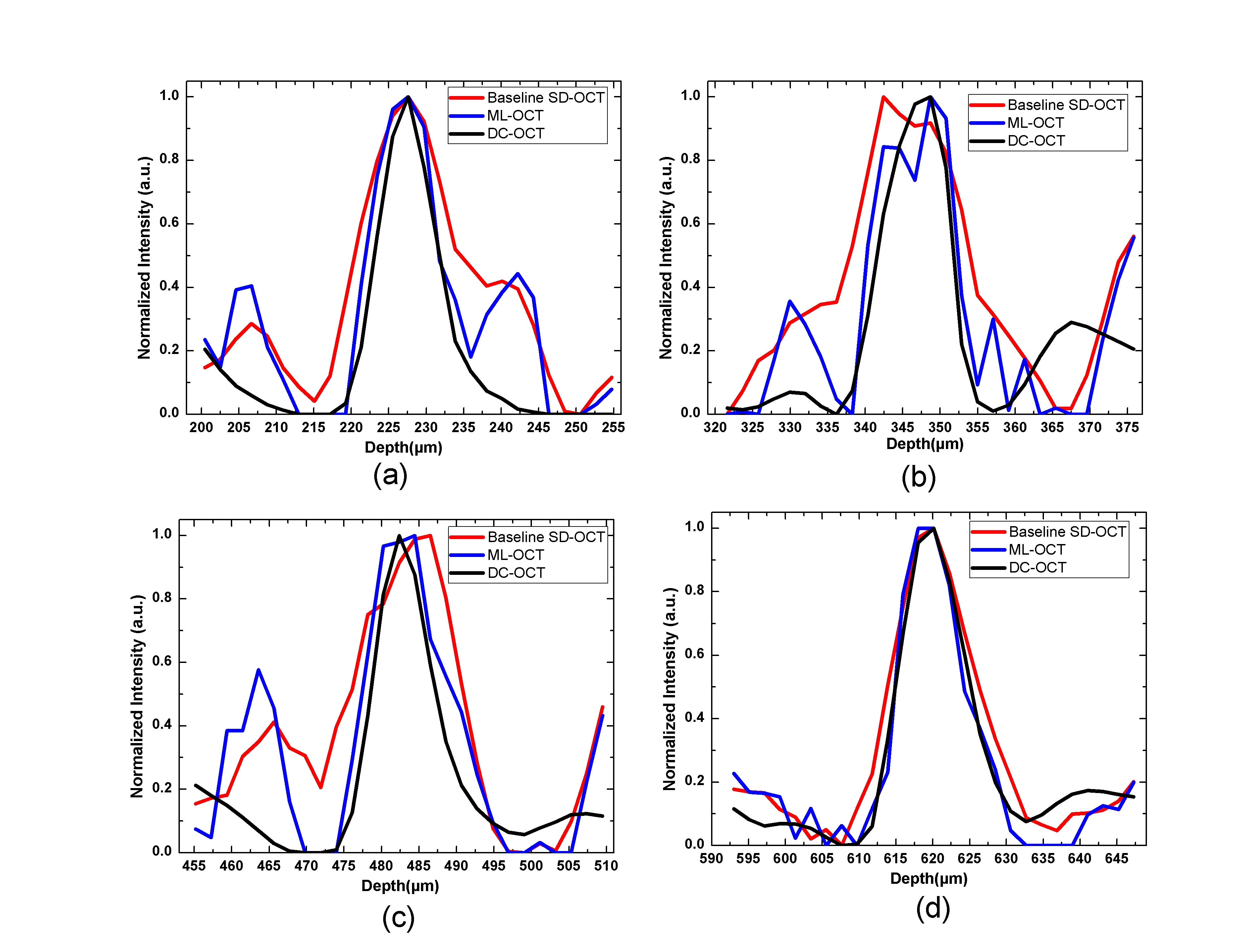}
	\caption{Calculated axial resolution PSFs for a specified axial line of grape data as marked in Figure 5 (a-l) using red color lines. In all subfigures, red color plots show the axial PSF of the baseline SD-OCT, blue color plots show the axial PSF of ML-OCT and black color plots represent the axial PSF of the proposed DC-OCT at different imaging depths.}

\end{figure}

In order to quantitatively evaluate the performance of our proposed DC-OCT system in terms of lateral and axial resolution enhancement as well as in compensating for the SNR drop-off, we performed imaging on a standard U.S. Air Force Test Chart (USAF-1951) resolution target. The resolution target was used for calculating the SNR curve, the lateral resolution curve, and the axial resolution curve of the baseline SD-OCT, ML-OCT, and proposed DC-OCT systems. The required experimental procedures for these calculations are fully explained in the \textbf{Methods} section. The calculated axial and lateral resolution curves are correspondingly displayed in Figure 2(a,b) and the SNR curves are plotted in Figure 2(c).

Calculated axial resolution curves in Figure 2(a) clearly show the effectiveness of the proposed DC-OCT system in terms of axial resolution enhancement along different imaging depths. Based on these plots, an average axial resolution improvement of about $2\mu m$ is obtained using the proposed DC-OCT system (compare red and black plots in Figure 2(a)) while the average axial resolution improvement is about $1.5\mu m$ when using the ML-OCT system (compare red and blue plots in Figure 2(a)).

According to the displayed lateral resolution curves in Figure 2(b), utilizing the proposed DC-OCT system results in a lateral resolution improvement of an average 0.6 $\mu m$ within all different imaging depths when compared to the baseline SD-OCT system, with the amount of improvement being even more at higher imaging depths ($>600 \mu m$). The comparison of blue and black curves in Figure 2(b) confirms the improved performance of the proposed DC-OCT system in terms of lateral resolution enhancement when compared to the tested ML-OCT system. Furthermore, the calculated lateral resolution curve for the tested ML-OCT system shows its limitation in lateral resolution enhancement at higher imaging depths ($>$400 $\mu m$), where the calculated lateral resolution for the ML-OCT system at these imaging depths drops below that of the baseline SD-OCT system.

The comparison of SNR curves in Figure 2(c) demonstrates an average SNR improvement of about $16 dB$ within different imaging depths using the proposed DC-OCT system when compared to the baseline SD-OCT system. Comparing the calculated SNR curves of the proposed DC-OCT system and the tested ML-OCT system (black and blue plots in Figure 2(c)) clearly proves the SNR improvements achieved by the proposed DC-OCT system when compared to the ML-OCT system at all imaging depths.  In fact, it can be observed that the proposed DC-OCT system can achieve an almost uniform SNR curve (black plot in figure 2(c)) across different imaging depths. Furthermore, the comparison of blue and red curves in Figure 2(c) illustrate the limitations of the tested ML-OCT system in dealing with the SNR drop-off, particularly at higher imaging depth ($>1100 \mu m$) where the SNR achieved using the ML-OCT system is lower than that of the baseline SD-OCT system.

For the case of biological samples (grape and lime), we perform a quantitative evaluation on the axial and lateral resolution enhancement at different imaging depths. To this aim, we calculate the axial and lateral PSFs (intensity plots) around chosen small cells nuclei or cell boundaries which are residing at different imaging depths of the biological sample. The PSFs are calculated by obtaining the intensity values at determined axial or lateral lines passing through a chosen nuclei or cell boundary. These lines are marked with red color in Figure 5(a-l) and in Figure 10(a-l) for the case of axial PSF calculation for the imaged grape and lime samples, and also are shown with red color lateral lines in Figure 7(a-l) and Figure 12(a-I) for the case of lateral PSF calculation for the imaged grape and lime samples. The calculated axial and lateral PSFs at different imaging depths are correspondingly displayed in Figure 6(a-d) and Figure 8(a-d) for the case of the grape sample and in Figure 11(a-d) and Figure 13(a-d) for the case of the lime sample. As we can see from the calculated axial and lateral PSFs of Figures 6,8 (a-d) as well as Figures 11,13 (a-d), utilizing the proposed DC-OCT system resulted in more narrower and smoother PSFs with less side lobe artifact when compared to the other tested systems, which shows the effectiveness of the proposed DC-OCT system in terms of compensating axial and lateral blurring in both axial and lateral directions and within different imaging depths while at the same time improving overall SNR.

%------Figure 7---------------------------
\begin{figure}[ht!]
	\centering
    \includegraphics[width=0.5\linewidth]{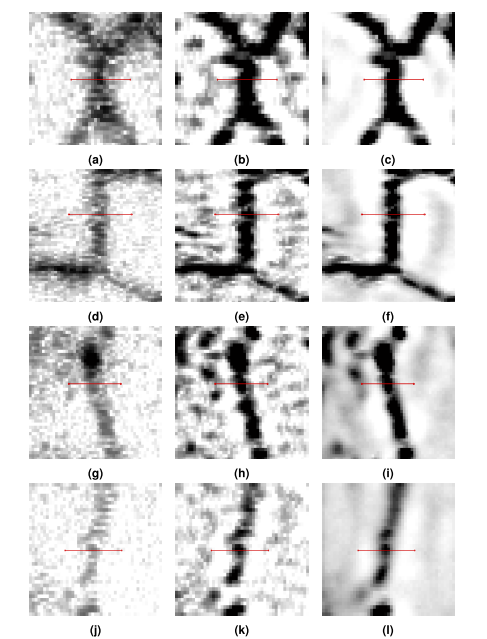}
	\caption{Enlarged regions representing different boundary cells of grape at various imaging depths as marked with green boxes in Figure 3(a) and for the baseline SD-OCT (a,d,g,j), ML-OCT (b,e,h,k) and proposed DC-OCT (c,f,i,l).}	
\end{figure}

%----Figure 8-------------------
\begin{figure}[ht]
	\centering
    \includegraphics[width=0.8\linewidth]{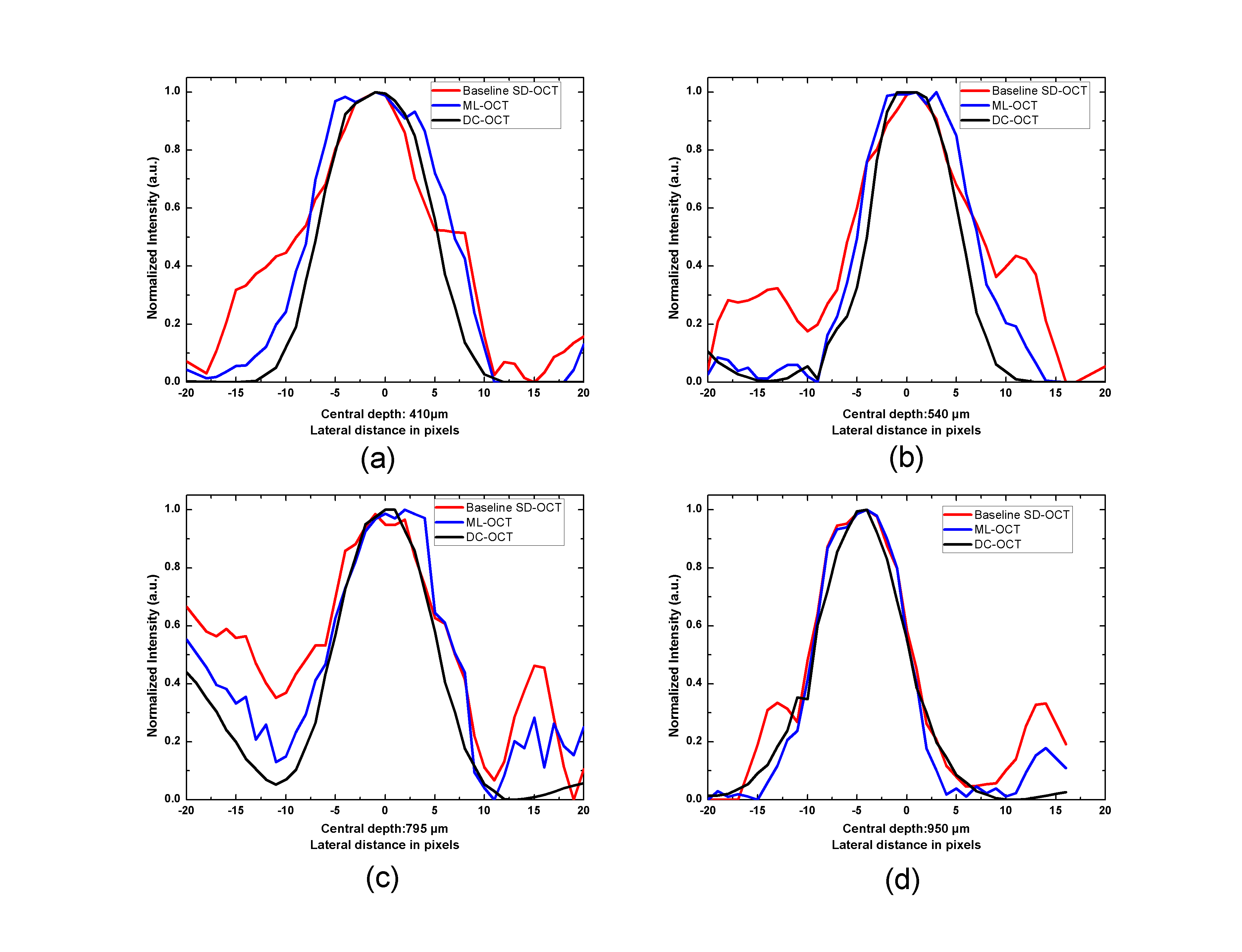}
	\caption{Calculated lateral resolution PSFs for a specified line of grape data as marked in Figure 7(a-l) using red color lateral lines. In all subfigures, red color plots shows the lateral PSF of the baseline SD-OCT, blue color plots show the lateral PSF of ML-OCT and black color plots represent the lateral PSF of the DC-OCT at central depths of (a) 410$\mu$, (b) 540$\mu$, (c) 795$\mu$ and (d) 950$\mu$.}
\end{figure}

%-------------Figure 9-----------------------
\begin{figure}[ht!]
	\centering
    \includegraphics[width=0.9\linewidth]{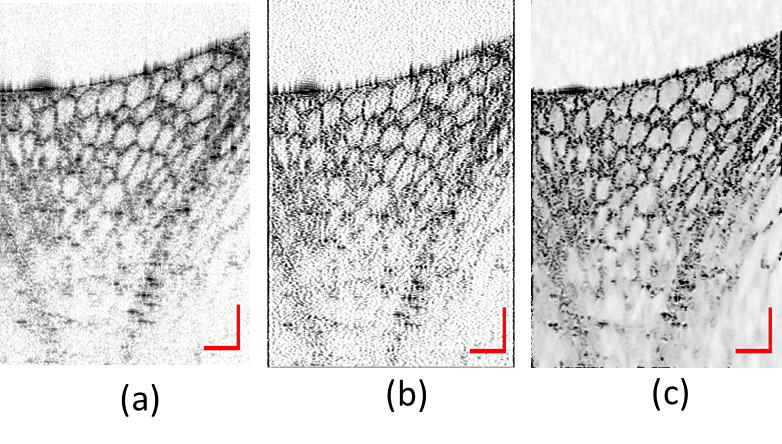}
	\caption{A slice of lime imaged using (a) baseline SD-OCT system, (b) ML-OCT system and (c) proposed DC-OCT system. The scale bars represent 80 $\mu m$ in both axial and lateral directions.}
	\
\end{figure}

\section*{Discussion}
In all SD-OCT imaging systems, there are several factors related to either the optical characteristics of the SD-OCT system or the different physical properties of the imaged sample that can often degrade the overall quality of resulting tomograms. These issues limit the ability to clearly visualize different micro-scale features and fine details of the imaged sample. Instead of tackling these issues with more complex and expensive optics solutions, we introduced a novel DC-OCT system that incorporates a Depth Compensating Digital Signal Processing (DC-DSP) module to more adaptively and cheaply overcome these limitations and achieve improved quality enhancement in the produced tomograms.  The DC-OCT system takes advantage of some inherent physical characteristics of the SD-OCT system such as the measured axial and lateral PSFs along with the calculated SNR curve of the system to digitally compensate for depth-varying axial and lateral resolution degradation, depth-varying SNR drop-off, and sidelobe artifact.

The integrated DC-DSP module is based on a unified MAP framework that incorporates a SF-CRF model to mitigate speckle noise and sidelobe artifact while improving the spatial resolution of the resulting tomograms and provide better visualization of the various fine structures and details of the imaged sample across different imaging depths. The quantitative results using both the standard USAF resolution target as well as samples of different biological tissues illustrate the effectiveness of the proposed DC-OCT system when compared to the state of the art ML-OCT system which is recognized for the depth-compensated SD-OCT imaging. While the ML-OCT system only compensates for the depth-dependent axial and lateral blurring~\cite{hojjatoleslami2013image}, the proposed DC-OCT system concurrently compensates for the effect of depth-varying axial and lateral degradation as well as the SNR drop-off along higher imaging depths and sidelobe artifact. The resulted resolution and contrast enhancements, along with improvements in SNR when using the proposed DC-OCT system helps in better visualization of small details and features of the biological samples such as cells nuclei and fine membranes of the cells and also is useful for increasing the accuracy of further required processing tasks such as cell segmentation and cell counting.

%-----------------------------Figure10----------------------------------------
\begin{figure}[ht!]
	\centering
    \includegraphics[width=0.5\linewidth]{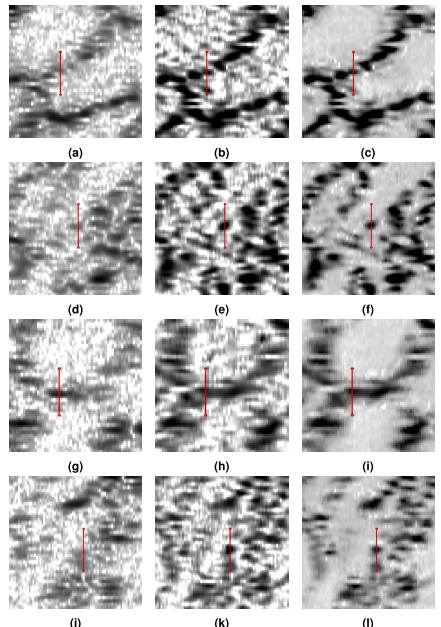}
	\caption{Enlarged regions representing different boundary cells and small details of lime at various imaging depths as marked with red boxes in Figure 3(b) and for the baseline SD-OCT (a,d,g,j), ML-OCT (b,e,h,k) and proposed DC-OCT (c,f,i,l).}	
\end{figure}

%------Figure 11--------------------------------
\begin{figure}[ht]
	\centering
    \includegraphics[width=0.8\linewidth]{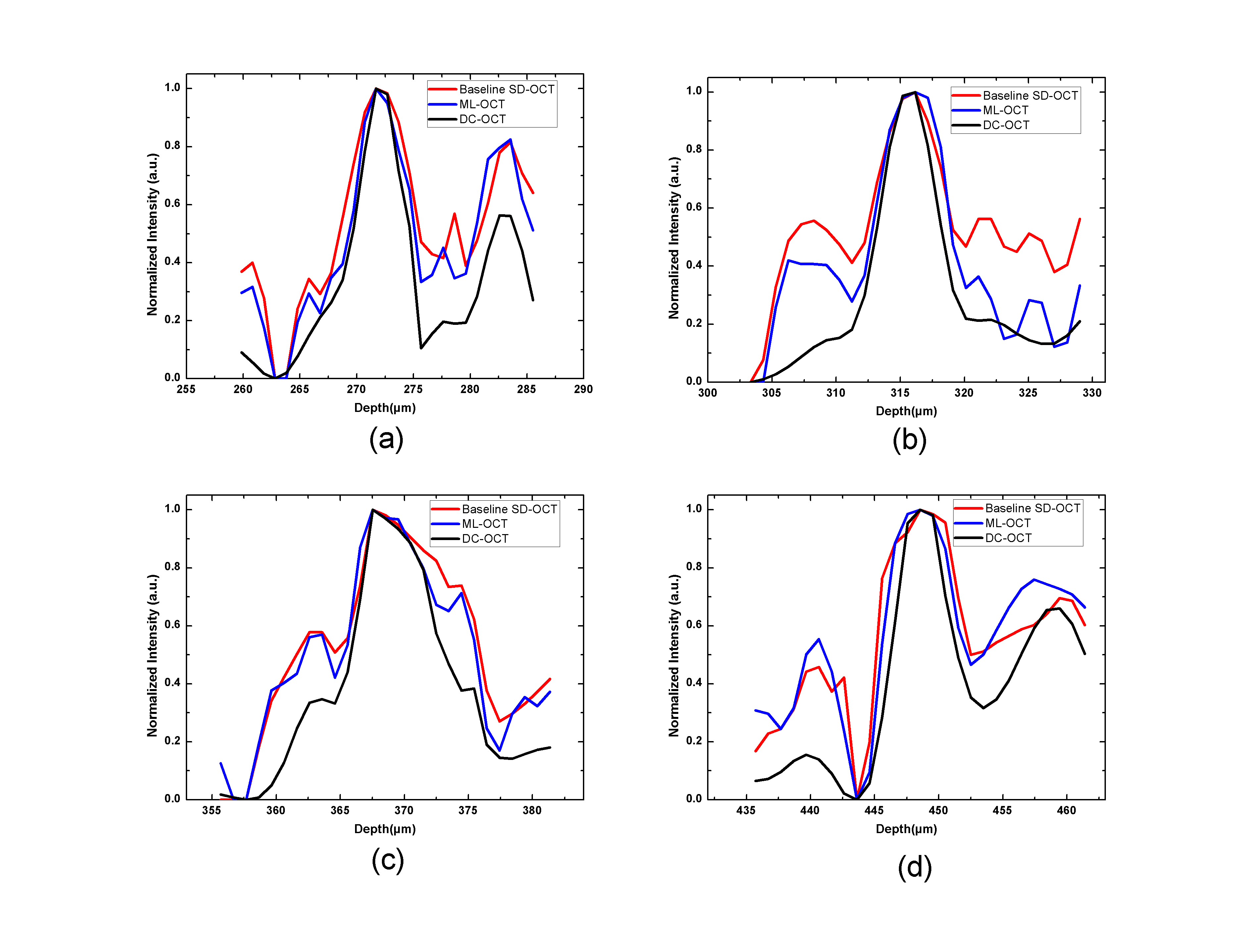}
	\caption{Calculated axial resolution PSFs for a specified axial line of lime data as marked in Figure 10 (a-l) using red color lines. In all subfigures, red color plots show the axial PSF of the baseline SD-OCT, blue color plots show the axial PSF of ML-OCT and black color plots represent the axial PSF of the proposed DC-OCT at different imaging depths.}
\end{figure}

%-------------------Figure 12------------------------------------------
\begin{figure}[ht!]
	\centering
    \includegraphics[width=0.5\linewidth]{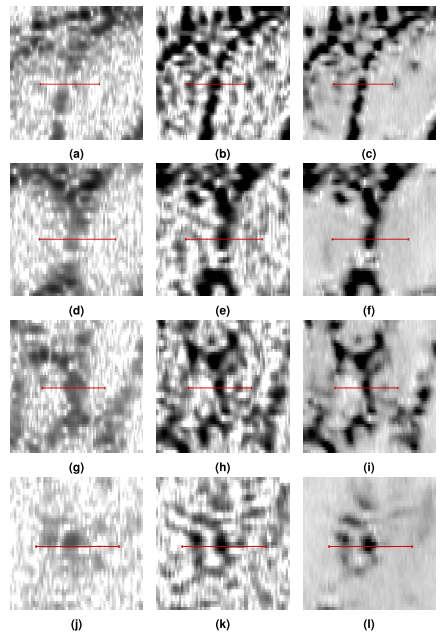}
	\caption{Enlarged regions representing different boundary cells and small details of lime at various imaging depths as marked with green boxes in Figure 3(b) and for the baseline SD-OCT (a,d,g,j), ML-OCT (b,e,h,k) and proposed DC-OCT (c,f,i,l).}	
\end{figure}

%----------figure13------------------------------------------------------------------------
\begin{figure}[ht]
	\centering
    \includegraphics[width=0.8\linewidth]{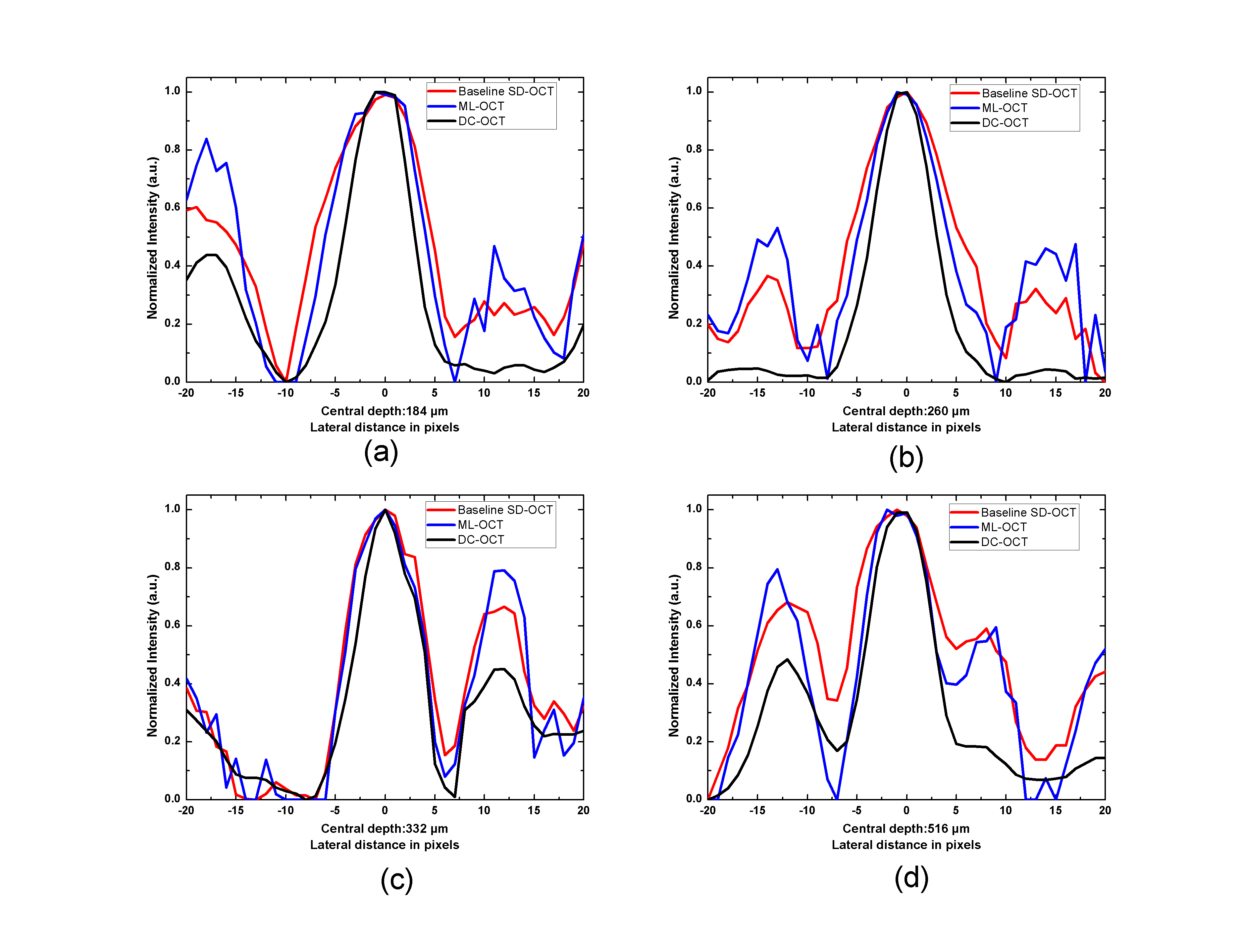}
	\caption{Calculated lateral resolution PSFs for a specified line of lime data as marked in Figure 12(a-l) using red color lateral lines. In all subfigures, red color plots shows the lateral PSF of the baseline SD-OCT, blue color plots show the lateral PSF of ML-OCT and black color plots represent the lateral PSF of the DC-OCT at central depths of (a) 184$\mu$, (b) 260$\mu$, (c) 332$\mu$ and (d) 516$\mu$.}
\end{figure}

\section*{Methods}

\subsection*{DC-OCT acquisition system setup}
In this study, we build upon a high-speed, ultrahigh resolution SD-OCT setup (Figure 1) for the proposed DC-OCT system to acquire tomograms of several biological samples such as grape and lime, as well as a standard U.S. Air Force Test Chart (USAF-1951) resolution target. The core of the utilized SD-OCT system is a fiber optic Michelson-Morley interferometer, interfaced to a broad bandwidth Super Luminescent Diode (SLD) as the light source (Superlum Ltd., with spectral range $\lambda c=1020$ nm, central bandwidth $\Delta \lambda=110 nm$ and output power $P_{out}=10 mW$). It utilizes a InGaAs camera (SUI, Goodrich Corp) with 1024 pixels to acquire interference fringes at 47kHz scanning speed.  The measurements made by the InGaAs camera, along with the axial and lateral PSF degradation functions and the measured SNR curve of the SD-OCT setup, are then used by the integrated DC-DSP module to digitally compensate for the degrading effects of the depth-dependent SNR roll-off, depth-varying axial and lateral resolutions as well as sidelobe artifact to produce an output depth-compensated tomogram.

\subsection*{Characterization of SD-OCT setup}
A prerequisite for obtaining depth-compensated tomograms using the proposed DC-OCT system is the calibration measurement and characterization of the depth-varying axial and lateral PSFs as well as the sensitivity fall-off curve (SNR curve) of the SD-OCT setup, which will act as prior information for the proposed integrated DC-DSP module.  To obtain the actual depth-varying axial PSF as well as the sensitivity fall-off curve of the utilized SD-OCT setup, a number of tomograms are acquired of an ideal reflection surface (here an uncoated mirror) as an object while the beam of sample arm is focused onto the mirror and the reference arm is transferred by 100 $\mu m$ at each step to calculate the PSF for each imaging depth. Next, a Gaussian function is fitted to each of the experimentally measured axial PSF points and then the value of Full Width at Half Maximum (FWHM) as well as the magnitude of these Gaussian functions are estimated for each specific depth. Finally, the measured depth-varying FWHM values are used to calculate the depth-varying axial PSF. The magnitude values of the estimated Gaussian function at different depths is used to generate the sensitivity fall-off curve.

Lateral resolution degradation is primarily happens due to the defocussing of the objective lens. During the imaging procedure, the converging beam focuses on the certain depths such that within the specific Rayleigh length, the degeneration of lateral resolution is minimal, but when beyond the Rayleigh length, the degeneration of lateral resolution gets larger. The depth-varying lateral PSF of the SD-OCT system is determined by the measurement of the FWHM of the lateral PSF via multiple acquisitions of the standard USAF resolution target using the baseline SD-OCT setup. To accomplish this, the beam is focused using 5X Mitutoyo (Edmund Optics,. Inc) microscopic objective lens, onto Group -1, Element 4 of the target, which has the width of $0.707 mm$ for each line pair. The edge of the line pair is placed at the center to get rid of any distortions introduced by the lateral degeneration. The lateral PSF is then defined as the mirror duplication of line pair edge slope and the lateral resolution is the FWHM of the slope. By fixing the reference mirror and shifting the target to different depth positions, the lateral FWHM at a range of depths is calculated.

\subsection*{Depth Compensating Digital Signal Processing (DC-DSP) module}

Given the measurements made by the InGaAs camera, along with the axial and lateral PSF degradation functions and the measured sensitivity fall off (SNR) curve of the SD-OCT setup obtained via the calibration characterization process, the integrated DC-DSP module digitally compensates for the degrading effects of the depth-dependent SNR roll-off, depth-varying axial and lateral resolutions as well as sidelobe artifact to produce an output depth-compensated tomogram. The underlying theory behind the DC-DSP module can be described as follows. Mathematically, the formation of the SD-OCT measurement, $M$, can be modeled as
\begin{align}
	M=F(V,S,H_{a}(z),H_{l}(z)).\xi(z),
\end{align}
where $\xi(\cdot)$ denotes the speckle noise that is inherent in OCT imaging, and $F(\cdot)$ defines a degradation function that relates the desired SD-OCT image $V$ to the SD-OCT measurement $M$. Therefore, the goal of the DC-DSP module is to obtain a depth-compensated tomogram $V$ given the SD-OCT measurement $M$. Ideally, the degradation function $F(\cdot)$ encompasses all the possible sources of image quality degradation either due to the effects of the SD-OCT imaging system or due to the structure of the imaged sample. In this study, as expressed in the formulation of equation (1), the degradation function $F(.)$ accounts for the axial PSF $H_{a}(z)$, lateral PSF $H_{l}(z)$, as well as the sidelobe artifact $S$, where $z$ denotes the imaging depth. The aforementioned mathematical model explicitly demonstrates the possibility of integrating the true physical characteristic curves of the utilized SD-OCT setup to acquire a depth-compensated image with improved overall image quality. To obtain the depth-compensated tomogram, $V$, we solve the following inverse problem:
\begin{equation}
	V=F^{-1}(M,S,H_{a}(z),H_{l}(z)).
\end{equation}
In the proposed integrated DC-DSP module, a unified Maximum a Posteriori (MAP) framework is employed to solve the inverse problem of equation (2), where the optimal estimate of the depth-compensated image (denoted by $\hat{V}$) is obtained by maximizing the posterior probability of $V$ given the acquired SD-OCT measurement $M$:
\begin{equation}
	\hat{V}=\underset{V^\prime}{argmax}P(V|M)
	\label{MAPformulation}
\end{equation}
where $P(V|M)$ denotes the posterior probability of $V$ given $M$. To better account and compensate for the presence of speckle noise and artifact in the measurement $M$, we incorporate a recently-developed random field model called the Stochastically Fully-connected Conditional Random Field (SF-CRF)~\cite{shafiee2014efficient,GSFCRF_2015_shafiee,DWI_2014_Shafiee} into the unified MAP framework of the DC-DSP module. Using the SF-CRF model, where $V$ and $M$ are represented as graphs with image pixels represented by nodes in the graphs, the posterior probability $P(V|M)$ can be expressed as
\begin{equation}
	P(V|M)=\frac{1}{\mathcal{Z}(M)}exp(-E(V,M)),
\end{equation}
where $E(\cdot)$ denotes an energy function defined over an assumed random field for both of the SD-OCT measurement $M$ and the depth-compensated image $V$
\begin{equation}
	E(V,M)=\sum_{i=1}^{n}\psi_{A}(v_{i},M)+\sum_{c\in C}\psi_{p}(v_{c},M),
	\label{eq:energy}
\end{equation}
where $i$ is the index number for the image pixels in its lexicographic representation form and $\mathcal{Z}(.)$ is a normalization term referred to as the partition function. Here, the energy function $E(\cdot)$ mainly encodes the direct relationships that exist between each pair of nodes $m_i$ and $v_i$ using a defined association potential function (unary term) $\psi_{A}$ and also models the interaction relationships between each pair of nodes $v_i$ and $v_j$ regarding to $M$ using the pairwise potential function $\psi_{p}$ in equation (5).

The logarithmic form of association energy function $\psi_{A}$ can be expressed by
\begin{equation}
	\psi_{A}(v_{i},M)=\frac{1}{\sigma\sqrt{2\pi}} exp\left(-\alpha\frac{\left(log(m_{i})-\sum_z log(F(v_{i},S,H_{a}(z),H_{l}(z)))\right)^{2}}{2\phi^{2}}\right).
	\label{eq:unary}
\end{equation}
The logarithmic domain is used to transform the multiplicative speckle noise inherent in SD-OCT measurements to an additive term which is common in general forward modeling. Here, $m_i$ denotes each single node in the graph representation for the SD-OCT measurement $M$ and $z$ denotes the imaging depth. The association energy function aims to obtain the best estimate $v_{i}$ for the depth-compensated image at each single node, $i$, and with respect to some of the SD-OCT measurement $M$ which are affected by the degradation function $F(.)$.

It is worth noting that due to the depth-dependent nature of the degradation function $F(.)$, here,  the overall effect of the degradation function on the acquired SD-OCT measurements $M$ is modeled using a non-homogeneous unary term.  Furthermore, to better preserve image boundaries and edges in the output depth-compensated image, the pairwise potential function $\psi_{p}(\cdot)$ is defined by the concept of stochastic cliques within a fully-connected random field, which can be expressed as follows:
\begin{equation}
	\psi_{p}(v_{c},M)=exp\left(\frac{\left\Vert N_{i}-N_{j}\right\Vert _{2}}{\sigma_c}\right)(v_{i}-v_{j})\,;\, c=\{i,j\},
\end{equation}
where $v_c$ is a subset of random variables $v_i$ and $v_j$ located at different axial and lateral directions and $\sigma_c $ expresses the standard deviation of the penalty function in the pairwise model for a specific clique structure $C$. Since the available measurements are non-homogeneous due to the overall effects of the degradation function $F(.)$, the effect of each node $v_j$ on node $v_i$ is modeled based on the dependency of the location node $v_j$ to the node $v_i$ in the defined lexicographic lattice and is defined for each pair of nodes separately. The penalty function is defined based on the similarity of each two nodes which is determined based on the SD-OCT measurements $M$. The set of measurements in specified neighborhoods $N_i$ and $N_j$ and centered at pixels $m_i$ and $m_j$ are utilized to compute the similarity of two nodes $v_i$ and $v_j$. Using the set of nodes instead of one node and computing the similarity makes the model more robust to the presence of speckle noise and artifact.

To account for the non-homogeneity of the SD-OCT measurements due to the depth-varying nature of the degradation function $F(.)$, we extend upon the stochastic clique structure proposed in~\cite{shafiee2014efficient} to incorporate the non-homogeneous relationships amongst the nodes into the model.  In this study, a specific stochastic clique is assigned to each pair of nodes in the defined stochastic graph based on the stochastic clique indicator function~\cite{shafiee2014efficient}, while the pairwise potentials among each pair of nodes are determined based on the connection that exist between each pair of depth-wise nodes.  Furthermore, the non-homogeneity of the clique structures is accounted for by utilizing different standard deviations in the penalty functions based on the imaging depth at which the nodes exist.  A detailed explanation of the graph representation used in the SF-CRF model is provided in the \textbf{Graph Representation} subsection. Based on the constructed $P(V|M)$, the final estimate of the output depth-compensated image ($\hat V$) is obtained by solving optimization problem of equation (3) using a gradient descent optimization technique, which will be described in the \textbf{Inference} subsection.\\

\textbf{Graph Representation.} The graph $G(\mathcal{V},\mathcal{E})$ represents the underlying structure of the random field model that is used in this study, where $\mathcal{V}$ is the set of nodes of the graph representing the states $V=\{v_i\}^{n}_{i=1}$ (here, these states are referring to the depth-compensated OCT image $V$) and $\mathcal{E}$ is the set of edges of the graph. Corresponding to each vertex in the graph $G(.)$ there is an observation $m_i\in M$ denotes to the SD-OCT measurements acquired using the baseline SD-OCT system. The defined graph $G$ is a fully-connected graph such that the pairwise clique structures which are incorporated in the inference are drawn stochastically. In this graph, the clique connectivity between each pair of nodes is determined based on a stochastic approach such that pairs of nodes that are closer in the defined random field should have higher probability to construct a clique structure than pairs of nodes that are farther apart in the defined random field. Furthermore, two similar nodes in the random field should have higher probability to be connected than two dissimilar nodes as it is encoded in the stochastic clique indicator~\cite{shafiee2014efficient}.

%----------figure14------------------------------------------------------------------------
\begin{figure}[ht]
	\centering
    \includegraphics[width=0.5\linewidth]{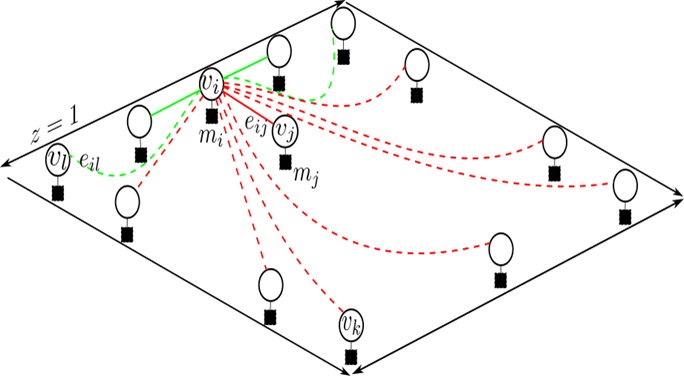}
	\caption{A realization of a stochastically fully connected conditional random field graph with non-homogeneous clique structure when $z$ represents the depth of imaging. A connectivity between two nodes is determined based on a distribution; each two nodes in the graph can be connected according to a probability drawn from this distribution. There is a SD-OCT measurement $m_i$ corresponding to each node $v_i$. The connectivity of each pair of two axial nodes ($v_i$ and $v_j$) and lateral nodes ($v_i$ and $v_l$) is distinguished by using different pairwise potentials $e_{ij}$ and $e_{il}$ due to the non-homogeneity of the OCT data. Closer nodes are connected with a higher probability (solid edges), whereas two nodes with a greater separation are less probable to be connected (dashed edges).}
\end{figure}

To explain these connections, let us consider Figure 14 where it is assumed that $v_i$ is in the first horizontal layer of the graph and each node can be connected to the all other nodes of the graph by some probability. According to our definition for the stochastically fully-connected graph, the probability of connectedness for the nodes pair $v_i$ and $v_j$ is higher than the probability of node pairs $v_i$ and $v_k$, since they have a smaller distance between each other. Furthermore, due to the data non-homogeneity, the defined pairwise potential related to the lateral direction is different from the axial one for each pair of nodes. As seen, the corresponding pairwise potential of two nodes $v_i$ and $v_l$ is different from the pairwise potential of $v_i$ and $v_j$ that makes the clique structures non-homogeneous. The reason for the non-homogeneity of the clique structure is that the degradation function $F(.)$ which is characterizing the degradation in the SD-OCT measurement varies with the depth and  therefore, the effective potential function changes based on the node's imaging depth $z$. According to the general definition of the CRFs~\cite{shotton2006textonboost}, all SD-OCT measurements can be incorporated into the unary and pairwise potential functions. Since the SD-OCT measurement, $M$, is acquired with some overlap between line scans, the combination of some measurements are utilized to estimate $v_i$ in the unary potential function.\\

\textbf{Inference.} Due to the large amount of measurements acquired by the SD-OCT system setup, regular inference methods are not applicable for solving the problem of obtaining depth-compensated tomograms from SD-OCT measurements as it has high computational complexity. Here, we address the computational complexity of the inference step by utilizing a gradient descent strategy~\cite{nocedal2006conjugate} to estimate the depth-compensated tomogram $V$. As discussed earlier, the problem of obtaining a depth-compensated tomogram $V$ given SD-OCT measurements $M$ is formulated in a MAP framework, with the optimal solution obtained by minimizing the energy function $E(\cdot)$ in~\eqref{eq:energy}:
\begin{align}
	V^\star = \underset{V'}{argmax} P(V|M) = \underset{V'}{argmin} E(V,M)
\end{align}
Here, the minimization of the energy function $E(\cdot)$ is done via a gradient descent strategy:
\begin{align}
	V^{t+1} = V^t - A \frac{ \nabla E(V,M)}{\nabla V}
\end{align}
where $V^{t+1}$ and $V^t$ are the solutions at iteration $t+1$ and $t$, $\frac{ \nabla E(V,M)}{\nabla V}$ is the energy gradient, and $A$ is the learning rate. The gradient descent strategy is applied on each node separately, therefore,
\begin{align}
	v_i^{t+1} = v_i^t  - \alpha \frac{ \partial \psi_{A}(v_i,M)}{\partial v_i} -  \beta \sum_{j  \in \mathcal{N}_i }\frac{\partial \psi_{p}(v_i,v_j,M)}{\partial v_i}
\end{align}
where  $\frac{ \partial \psi_{A}(v_i,M)}{\partial v_i}$ and $\frac{\partial \psi_{p}(v_i,v_j,M)}{\partial v_i}$ represent the gradient of unary and pairwise potentials in a specified neighbors $\mathcal{N}_i$ defined for each node $v_i$. In this formulation, $A = \{\alpha, \beta\}$ respectively denote to the data-driven learning rate and learning rate with respect to the spatial information. Assigning two different learning rates facilitates the procedure of adjusting the effect of each factor (i.e., data importance or spatial consistency) based on the application. For example, the weight of the unary potential, $\alpha$, should be higher at the conditions where the acquired tomograms exhibit high SNR, while weight of $\alpha$ should be lower for the cases in which the acquired tomograms exhibit lower SNR. % \hl {We can talk more here!!}

\section*{Implementation}
The algorithms pertaining to the unified framework in the integrated DC-DSP module are implemented in embedded C++ code in MATLAB (The MathWorks, Inc.) and tested on an AMD Athlon II X3 3.10 GHz machine with 12GB of RAM. Comprehensive parametric analysis was performed to obtain the optimal parameters for the proposed DC-OCT system as well as the tested ML-OCT~\cite{hojjatoleslami2013image} system, which is also capable of producing depth-compensated tomograms. The only free parameters for generating a depth-compensated tomogram using the proposed DC-OCT system are $\alpha$, $\beta$ and the number of required iterations for performing the digital reconstruction. By performing an empirical test, these parameters were set as $\alpha=1$, $\beta=0.3$ and 30 iterations for the case of testing standard USAF resolution target while they were set as $\alpha=1$, $\beta=0.2$ and 30 iterations for the case of generating depth-compensated tomograms of grape and lime samples using the proposed DC-OCT system. For the case of generating depth-compensated tomogram using the tested ML-OCT system, the only free parameter is the number of required iterations for reconstructing a depth-compensated OCT tomogram. The number of iterations was set to 30 when the tested ML-OCT system was used for generating depth-compensated SD-OCT tomograms from the standard USAF resolution target as well as grape and lime samples. The computation times for obtaining a depth-compensated tomogram using the proposed DC-OCT system as well as for the tested ML-OCT system were similar and were both typically in the range of a few seconds.
\bibliography{sample}

\section*{Acknowledgements}
This work was supported by the Natural Sciences and Engineering Research Council of Canada, Canada Research Chairs Program, and the Ontario Ministry of Research and Innovation.

\section*{Author contributions statement}
B.T. and K.B. designed the baseline SD-OCT system setup and characterized the physical properties of the SD-OCT system. A.B. and A.W. designed the DC-DSP module and the associated algorithms. B.T. performed the image acquisitions. A.B. analyzed the data and performed the performance analysis. M.S. was involved in the integration of stochastically fully connected conditional random field and helped in problem formulation. All authors contributed to writing the paper and to the editing of the paper.

%\section*{Figure Legends}

\end{document}